\documentclass[%
prl,
twocolumn,
amsmath,
amssymb,
reprint,
superscriptaddress,
footinbib,
 amsmath,amssymb,
 aps,
]{revtex4-1}

\usepackage{graphicx}
\usepackage{dcolumn}
\usepackage{bm}
\usepackage{natbib}
\usepackage{color}
\usepackage{soul}
\usepackage{changes}

\begin{document}
\preprint{APS/123-QED}
\title{Spectral Purification of Microwave Signals with Disciplined Dissipative Kerr Solitons}

\author{Wenle Weng}
\thanks{W. Weng and E. Lucas contributed equally to this work.}
\affiliation{{\'E}cole Polytechnique F{\'e}d{\'e}rale de Lausanne (EPFL), CH-1015 Lausanne, Switzerland}

\author{Erwan Lucas}
\thanks{W. Weng and E. Lucas contributed equally to this work.}
\affiliation{{\'E}cole Polytechnique F{\'e}d{\'e}rale de Lausanne (EPFL), CH-1015 Lausanne, Switzerland}

\author{Grigory Lihachev}
\affiliation{Russian Quantum Center, Skolkovo 143025, Russia}
\affiliation{Faculty of Physics, M.V. Lomonosov Moscow State University, 119991 Moscow, Russia}

\author{Valery E. Lobanov}
\affiliation{Russian Quantum Center, Skolkovo 143025, Russia}

\author{Hairun Guo}
\affiliation{{\'E}cole Polytechnique F{\'e}d{\'e}rale de Lausanne (EPFL), CH-1015 Lausanne, Switzerland}

\author{Michael L. Gorodetsky}
\affiliation{Russian Quantum Center, Skolkovo 143025, Russia}
\affiliation{Faculty of Physics, M.V. Lomonosov Moscow State University, 119991 Moscow, Russia}

\author{Tobias J. Kippenberg}
\email{tobias.kippenberg@epfl.ch}
\affiliation{{\'E}cole Polytechnique F{\'e}d{\'e}rale de Lausanne (EPFL), CH-1015 Lausanne, Switzerland}

\date{\today}

\begin{abstract}
Continuous-wave-driven Kerr nonlinear microresonators give rise to self-organization in terms of dissipative Kerr solitons, which constitute optical frequency combs that can be used to generate low-noise microwave signals. Here, by applying either amplitude or phase modulation to the driving laser we create an intracavity potential trap to discipline the repetition rate of the solitons. We demonstrate that this effect gives rise to a novel spectral purification mechanism of the external microwave signal frequency, leading to reduced phase noise of the output signal. We experimentally observe that the microwave signal generated from disciplined solitons is injection-locked by the external drive at long time scales, but exhibits an unexpected suppression of the fast timing jitter. Counter-intuitively, this filtering takes place for frequencies that are substantially lower than the cavity decay rate. As a result, while the long-time-scale stability of the Kerr frequency comb's repetition rate is improved by more than 4 orders of magnitude, the purified microwave signal shows a reduction of the phase noise by 30~dB at offset frequencies above 10~kHz.
\end{abstract}
\pacs{Valid PACS appear here}
\maketitle
\textit{Introduction.}---Low-noise microwave signals play a vital role in a wide range of industrial and scientific applications, including telecommunication networks \cite{pozar2001microwave}, radar/LIDAR systems \cite{maleki2011sources} as well as in fundamental research such as long baseline interferometry \cite{grop2010elisa} and tests of fundamental constants \cite{stanwix2005test, nagel2015direct}. Traditionally, the microwave signals with the best spectral purity were provided by cryogenic microwave oscillators \cite{grop201010, hartnett2012ultra}. Owing to the advancement of mode-locked-laser frequency combs and optoelectronics, new photonic-based ways of generating ultralow-noise microwaves have been proposed and demonstrated, such as optical frequency division \cite{fortier2011generation,Quinlan:2013aa, xie2017photonic}, electro-optical frequency division \cite{li2014electro}, or Brillouin lasing in microresonators \cite{li2013microwave, loh2016microrod}.

Recently, dissipative Kerr solitons (DKS) in optical microresonators \cite{herr2014temporal,akhmediev2005dissipative} have been attracting surging interests thanks to their self-organizing mechanism that results from the double-balance between nonlinearity and anomalous dispersion, as well as between parametric gain and cavity loss. DKS offer high coherence, broad bandwidth and microwave-repetition rate frequency combs (also referred to as soliton microcombs \cite{kippenberg2018dissipative}), and have been applied successfully to ultrafast ranging \cite{trocha2018ultrafast, suh2018soliton}, dual-comb spectroscopy \cite{suh2016microresonator, pavlov2017soliton,dutt2018chip},  calibrating astrophysical spectrometer \cite{Obrzud:2019aa,Suh:2019aa}, as well as optical frequency synthesis \cite{spencer2018optical}. Like mode-locked-laser frequency combs, soliton microcombs can function as a frequency link between the microwave/radio-frequency (RF) domain and the optical domain \cite{papp2014microresonator, del2016phase}. In particular, microcomb-based microwave oscillators hold great promise of providing a robust, portable and power-efficient way to synthesize pure microwave tones \cite{liang2015high}. In contrast to microresonator-based approaches of generating microwave signals using Brillouin lasers, the frequency of the generated signal is mainly determined by the cavity free spectral range (FSR), rather than the host material property of the resonator, thus offering control over the microwave center frequency. However, this flexibility comes at a price: reaching a good longterm stability requires the ability to control the comb repetition rate ($f_\text{rep}$) and the carrier-envelope offset ($f_\text{ceo}$) and discipline them to optical references or RF clocks. To obtain such ability most previous efforts focused on using active feedback to correct thermal drifts and noises \cite{jost2015all, huang2016broadband} and utilizing sophisticated structure design for appropriate actuation \cite{papp2013mechanical, del2016phase, lim2017chasing}. 

\begin{figure} [t]
\centerline{\includegraphics[width=\columnwidth]{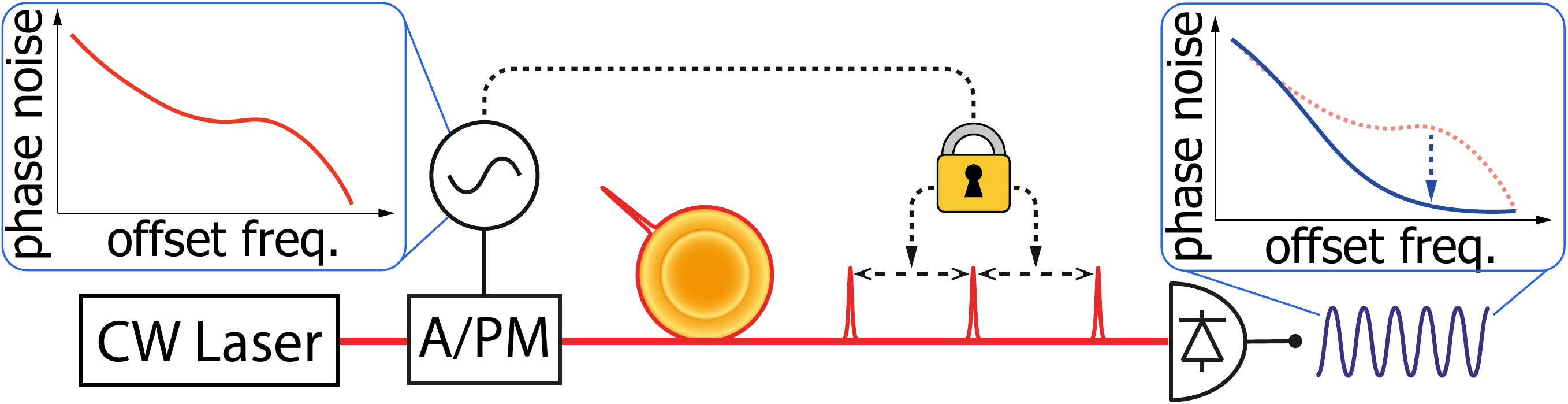}}
\caption{The concept of a microcomb-based microwave spectral purifier. A commercially-available electronic microwave oscillator (Rohde\&Schwarz SMB100A) is used to modulate the pump laser, leading to the injection locking of solitons, thus providing a long-term frequency reference to the soliton repetition rate. The generated microwave exhibits a reduced phase noise level due to the nonlinear soliton dynamics, leading to noise reduction of the microwave signal for Fourier frequencies far away from the carrier.}
\label{fig1}
\end{figure}

In this work, we use DKS in a crystalline microresonator to purify a 14.09\,GHz microwave signal. The phase noise of the purified signal approaches -130\,dBc/Hz at 10\,kHz offset frequency, which is at the level achieved by the state-of-the-art microresonator-based optoelectronic oscillators and the previously reported best results obtained with undisciplined DKS and narrowband RF filter \cite{maleki2011sources,liang2015high}. We adapt the microwave injection-locking technique that was previously used to stabilize modulation-instability (MI) combs \cite{papp2013parametric, papp2014microresonator} to discipline the soliton stream by creating intracavity potential gradient that traps the solitons. This mechanism not only relies on linear cavity filtering, but exploits further the dynamics of DKS, and allows to reduce substantially the phase noise of an external microwave drive. Owing to the dynamical attractor of the soliton state, the stability of the disciplined solitons exhibits strong robustness against incoherent perturbations contained in the injected signals \cite{leo2010temporal}, thus efficiently dissipating noises in a coherent system. This self-purifying mechanism leads to the reduction of the injected microwave phase noise, allowing the nonlinear cavity in the soliton state to act as a passive spectral purifier that can improve the performance of an external off-the-shelf electronic oscillators. As depicted in Fig.~\ref{fig1}, the disciplined-DKS-based microwave purifier constitutes in itself a $f_{\text{rep}}$-stabilized frequency comb and a spectrally pure microwave generator into a single device.

\begin{figure*} [t]
\centerline{\includegraphics[width=2\columnwidth]{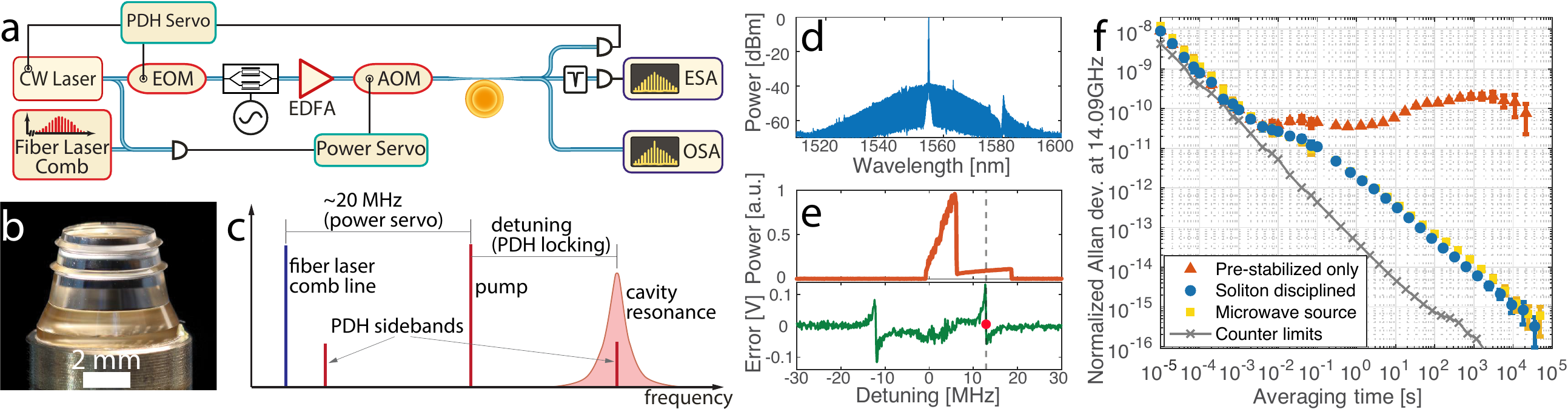}}
\caption{(a) The experimental setup. (b) The MgF$_2$ resonator used in the experiment. (c) Illustration of the PDH offset locking and the pre-stabilization scheme. (d) Optical spectrum of the soliton microcomb. (e) Generated comb power as the laser is scanned across the pumped resonance (upper) and the corresponding PDH error signal (lower). The red dot indicates the locking point. (f) Allan deviations of $f_{\text{rep}}$ when the Kerr comb is pre-stabilized and DKS-disciplined respectively. We counted $f_{\text{rep}}$ with a $\Pi$-type frequency counter that is referenced to the same frequency source (relative frequency instability $<1\times10^{-12}$ at 1\,s averaging time) to which $f_{\text{mod}}$ is referenced.}\label{fig2}
\end{figure*}

\textit{Experiment.}---The experimental setup is shown in Fig.~\ref{fig2}\,(a). A 1555-nm laser is amplified by an Erbium-doped fiber amplifier (EDFA) and 200\,mW optical power is coupled into a z-cut magnesium fluoride (MgF$_2$) whispering-gallery-mode resonator with a FSR of $14.09$\,GHz via a tapered fiber. A single-soliton-state DKS comb is generated by scanning the laser over a resonance with a loaded quality-factor ($Q$) of $1.3\times10^9$ to reach the step-like range where solitons are formed \cite{herr2014temporal}. To stabilize the effective laser detuning with respect to the cavity resonance, we apply phase modulation to the laser with an electro-optic modulator (EOM) to generate Pound-Drever-Hall (PDH) error signals. The laser frequency is locked to the high-frequency PDH sideband by setting the lock point of the servo to the center of the sideband resonance which is indicated in Fig.~\ref{fig2}\,(e). The frequency of the laser is then compared with a tooth of a stabilized fiber-laser-based comb, and the frequency difference is stabilized at 20\,MHz through a slow thermal actuation on the cavity by active control of the pump power with an acousto-optic modulator (AOM). As illustrated in Fig.~\ref{fig2}\,(c), with the two servos this ``pre-stabilization" scheme stabilizes both the pump laser frequency and the pump-cavity detuning. As a result, the stability of $f_\text{rep}$ is improved by up to 2 orders of magnitude at time scales of $>10$\,s (see Fig.~\ref{fig2}\,(f)), allowing the time-consuming measurement of phase noise via cross correlation to be carried out properly. One should note that the fiber-laser-based comb can be replaced with a laser stabilized by a reference cavity \cite{PhysRevLett.118.263202} or an atomic vapor cell \cite{liang2017stabilized}, and that with improved thermal isolation \cite{liang2015high} or self-referenced stabilization \cite{weng2015stabilization} the entire setup can be more compact.

\begin{figure*} [t]
\centerline{\includegraphics[width=2\columnwidth]{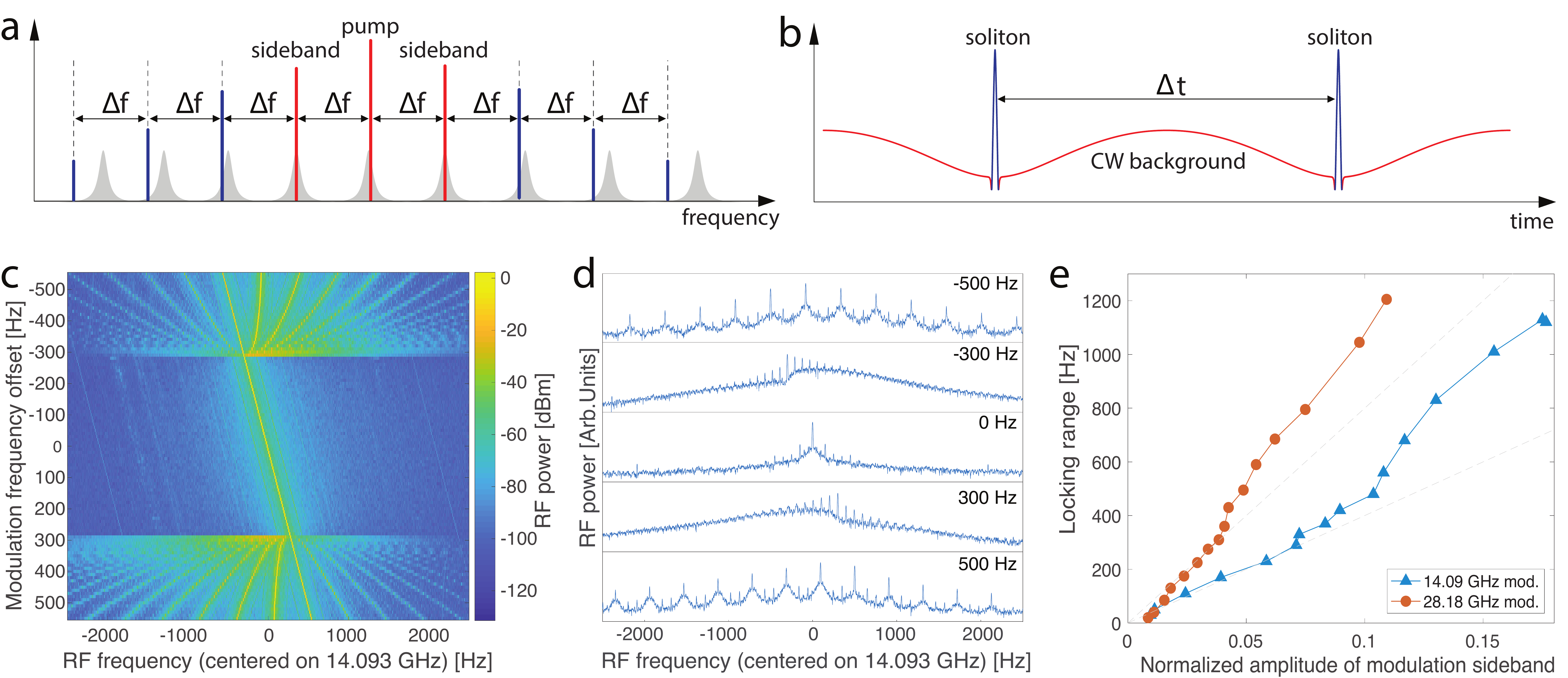}}
\caption{(a) In frequency domain, the difference between the pump laser and the modulated sidebands sets the microcomb $f_{\text{rep}}$. (b) In time domain, the potential of the modulated CW field traps solitons, thus locking $f_{\text{rep}}$ of the soliton train. (c) Evolution of the spectrum around 14.09\,GHz. (d) Snapshots with different $f_\text{mod}$ detuned from free-running $f_{\text{rep}}$ (indicated in the upper-right corners) showing the typical states of unlocked (-500\,Hz and 500\,Hz), quasi-locked (-300\,Hz and 300\,Hz) and injection-locked (0\,Hz). (e) Locking ranges with varied AM strengths for modulation frequencies around $f_\text{rep}$ (blue circles) and $2\times f_\text{rep}$ (red circles) respectively.}
\label{fig3}
\end{figure*}

The injection locking of the soliton repetition rate is implemented by applying amplitude modulation (AM) or phase modulation (PM) on the pump laser, at a frequency close to the FSR. Intuitive illustrations of how the injection locking works are presented in Fig.~\ref{fig3}\,(a) and (b). From a frequency domain perspective, the modulation frequency defines $f_{\text{rep}}$ through parametric four-wave-mixing. In the time domain, a modulated CW field traps solitons and disciplines $f_{\text{rep}}$ correspondingly. In this proof-of-principle experiment we use a synthesizer to drive the AM/PM modulator but the input microwave signal could be derived from a frequency-multiplied clock oscillator or a voltage-controlled oscillator (VCO). The modulation frequency $f_{\text{mod}}$ is swept around the free-running $f_{\text{rep}}$ ($\sim14.09$\,GHz) and we observe that $f_{\text{rep}}$ is injection-locked by the input microwave signal. Fig.~\ref{fig3}\,(c) shows the evolution of the microwave spectrum of the DKS as we slowly swept the AM $f_{\text{mod}}$. When the difference between $f_{\text{mod}}$ and the free-running $f_{\text{rep}}$ is larger than $\sim400$\,Hz, multiple spectral components including $f_\text{mod}$ (the strongest), $f_\text{rep}$ (the second strongest) and multiple harmonics are observed in the spectra, indicating an absence of injection locking. As $f_{\text{mod}}$ is approaching the free-running $f_{\text{rep}}$, the spectrum displays typical frequency-pulling effect as $f_\text{rep}$ is pulled towards $f_\text{mod}$ \cite{razavi2004study}. When the difference between $f_{\text{mod}}$ and free-running $f_{\text{rep}}$ is less than $\sim300$\,Hz all the spectral components merge into a major one, indicating that the $f_{\text{rep}}$ is synchronized to $f_{\text{mod}}$, i.e. the soliton stream is locked to the external drive. We measured the frequency instabilities of the injected-locked $f_{\text{rep}}$ against $f_{\text{mod}}$, which is also presented in Fig.~\ref{fig2}\,(f). The Allan deviation shows that at time scales of $>0.1$\,s the fluctuations of $f_{\text{rep}}$ has been suppressed significantly -- up to more than 4 orders of magnitude at averaging time of 1000\,s, indicating that the disciplined DKS tightly follow the injected microwave frequency.

We acquire the locking range from the evolution of the RF spectrum, and repeat the measurement with varied modulation strength. As shown in Fig.~\ref{fig3}\,(e), with the normalized amplitude of the modulation sideband below 0.07, the locking range rises monotonically with almost perfect linearity as the modulation strength increases. With stronger modulation the slope of the locking range scaling increases, which is attributed to the appearance of higher-order modulation sidebands that increase the gradient of the potential and trap the solitons more effectively \cite{taheri2017optical,taheri2015soliton,jang2015temporal,lobanov2015generation}. For the same reason, we observe that the locking range increases by nearly a factor of 2 when we measure the locking range with $f_\text{mod}$ around $2\times f_\text{rep}$ ($\sim28.18$\,GHz).

\begin{figure} [htb]
\centerline{\includegraphics[width=0.95\columnwidth]{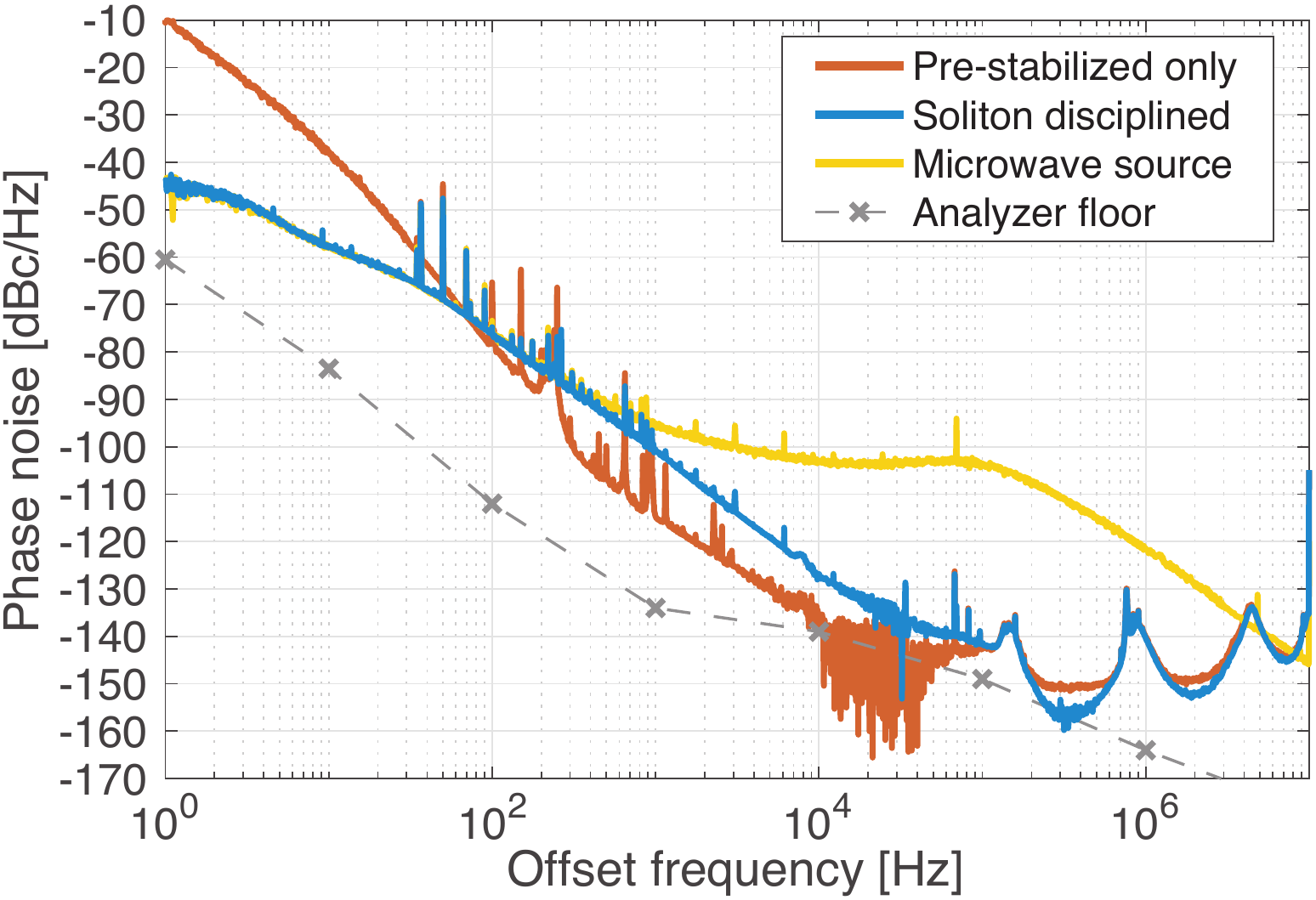}}
\caption{Phase noise spectra of the soliton repetition rate with and without PM injection locking. The phase noise of the input microwave signal is also presented, showing that the injection locking reduces the noise level by nearly 40\,dB for offsets at 100\,kHz. The crosses and the dashed line show the noise floor of the phase noise analyzer.}
\label{fig6}
\end{figure}

\textit{Spectral purification effect.}---To characterize the spectral purity at $f_{\text{rep}}$, the out-coupled soliton stream is filtered by fiber Bragg grating filters (FBG) to suppress the pump light and then amplified by an EDFA and subsequently attenuated to $\sim5$\,mW before being registered by a fast photodetector. We use a phase noise analyzer to measure the phase noise of the 14.09\,GHz signal output by the photodetector. Fig.~\ref{fig6} presents the single-sideband (SSB) phase noise level when PM injection locking was performed. One should note that very similar results were also observed with AM injection locking. At low offset frequencies between 1 -- 100\,Hz the injection-locked DKS show improved noise level that is in agreement with the phase noise level of the input RF tone due to the better long-term frequency stability provided by the injected microwave signal, as confirmed by the Allan deviations. This result shows that the soliton stream is strictly disciplined by the potential trap at low frequency ranges. Remarkably, at offset above 100\,Hz the spectrum of the injection-locked $f_{\text{rep}}$ mostly maintains the intrinsic high quality, which is several orders of magnitude lower than the input microwave in terms of phase noise level. We note that this purifying effect cannot be explained by the cavity filtering since the frequency range where the purification is observed is $\sim3$ orders of magnitude lower than the loaded cavity resonance bandwidth ($\sim150$\,kHz).  At offset frequencies above 30\,kHz a reduction of the input microwave phase noise level by 30\,dB is achieved, showing the exceptional spectral purifying ability of the disciplined DKS. 

\textit{Simulation of soliton spectral purification.}---In order to study the mechanism of the observed spectral purification, we performed simulations of PM-to-PM transfer function based on the Lugiato-Lefever equation (LLE) \cite{lugiato1987spatial}. The model is similar to the one described in \cite{guo2017intermode}, which is expressed as:
\begin{multline}
\label{eq_lle_like}
\frac{\partial {\tilde A}_{\mu}(t)}{\partial t} = \left( { - \frac{\kappa}{2} + i(2\pi\delta) + i D_{\rm int}(\mu) } \right){\tilde A}_{\mu} \\
- i g {\mathcal F} \left[ {|A|^2 A} \right]_{\mu} + {\sqrt{\kappa _{\rm ex}} s_{\rm in} (\delta '_{\mu 0} + \delta '_{\mu \pm 1} i \frac{\epsilon}{2} e^{\pm i \Omega t})}
\end{multline}
where ${{\tilde A}_{\mu}}$ and ${{A}}$ are the spectral and temporal envelopes of DKS respectively (related via ${A(t)=\sum_{\mu}{\tilde A}_{\mu} e^{-i \mu D_1 t}}$), ${\kappa}$ is the cavity loss rate, ${g}$ is the single photon induced Kerr frequency shift, ${\kappa_{\rm ex}}$ is the coupling rate and ${|s_{\rm in}|^2}$ denotes the power of the laser pumping the central mode, ${\delta '_{\mu 0/\pm1}}$ is the Kronecker delta, and ${{\mathcal F}[~]_{\mu}}$ represents the $\mu$'th frequency component of the Fourier series. We include third oder dispersion in $D_{\rm int}(\mu)$. A pair of PM sidebands are included in the last term of the equation, where $\epsilon$ indicates the amplitude of the modulation sidebands, and $\Omega$ is the frequency difference between the FSR and the input microwave signal.

\begin{figure} [t]
\centerline{\includegraphics[width=0.95\columnwidth]{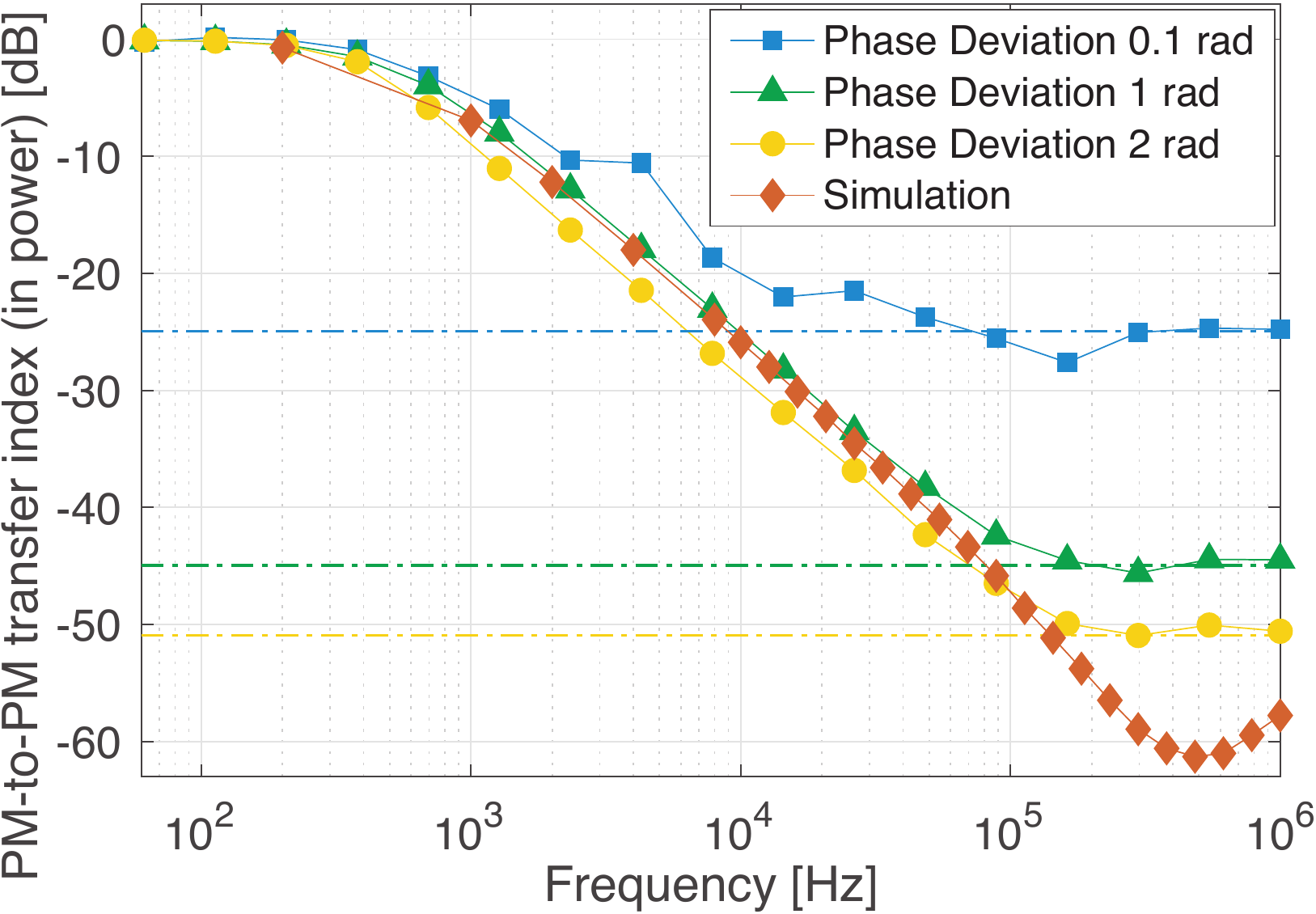}}
\caption{Experimentally measured and numerically simulated phase noise transfer indices for the PM-to-PM noise transfer from injected microwave signal to $f_{\text{rep}}$ of the DKS stream. The flat floors of the experimental data at frequencies above 100\,kHz are due to the noise floor of the phase noise analyzer during the IQ measurement, which are indicated by the dash-dot lines.}
\label{fig8}
\end{figure}

Adapting the technique used in \cite{matsko2015noise}, we introduce phase modulation on the microwave signal with phase deviation of 0.1 radian and varied modulation frequencies from 200\,Hz to 1\,MHz. The phases of the purified microwave signal can be derived from the comb spectra with
\begin{equation}
\label{phase}
\Psi (t) = \text{Arg} \left[ e^{i \omega_{\rm in} t} \sum_{\mu} \tilde A_{\mu} \tilde A_{\mu-1}^{\dagger} \right]
\end{equation}
where $\omega_{\text{in}}$ is the frequency of the input microwave signal. We use pump power of 200\,mW and $\epsilon=0.32$ for the numerical simulation \cite{SI}. The results are presented in Fig.~\ref{fig8}. The simulated transfer function follows a typical first-order lowpass filtering effect, showing a magnitude that is close to unity at low frequency (200\,Hz). For higher offset frequencies the magnitude decreases with a slope of -20\,dB/decade, reaching a minimum of $\sim-63$\,dB around 500\,kHz, thus revealing a significant phase noise suppression in the soliton state. To verify the simulated results, we apply PM with varied phase deviation on the injected microwave signal and record the resulting phase deviation on the soliton repetition rate with an in-phase-and-quadrature (IQ) demodulator \cite{SI}. The experimentally measured transfer functions are plotted in the same figure. From the comparison we see that at low frequencies the experimental results and the simulation are in satisfactory agreement. However, at frequencies above $\sim100\,$kHz the experimental curves show flat floors, which are attributed to the detection noise floor introduced by the analyzer we use to perform the measurement. This instrumental noise floor is confirmed by increasing the modulation strength, which improves the dynamic range of our measurement.

\textit{Conclusion.}---We have experimentally and numerically demonstrated a novel phase noise purifying mechanism by disciplining dissipative solitons with potential traps. The comb repetition rate drift, which is a major limitation in microcavities, was thereby suppressed, while this parameter was stabilized to a reference oscillator. The high frequency noise of the trapping signal was self-purified, at frequency offsets well below the cavity resonance bandwidth. Our technique reveals the unique dynamical stability of the self-organized temporal solitons. The exceptional phase noise level achieved with the purified microwaves shows that disciplined DKS are competitive with other state-of-the-art optical-microresonator-based microwave oscillators in terms of generating low-noise microwave signals with miniaturized device. It could also facilitate the application of microcombs in coherently averaged dual-comb spectroscopy \cite{coddington2016dual} and coherent optical telecommunication \cite{marin2017microresonator}.

This publication was supported by Contract D18AC00032 (Driven and Nonequilibrium Quantum Systems, DRINQS) from the Defense Advanced Research Projects Agency (DARPA), Defense Sciences Office (DSO) and funding from the Swiss National Science Foundation under grant agreement no.\,163864 and no.\,165933, and by the Russian Foundation for Basic Research under project 17-02-00522. W.W. acknowledges support by funding from the European Union's Horizon 2020 research and innovation programme under Marie Sklodowska-Curie IF grant agreement no.\,753749 (SOLISYNTH).


\bibliography{Ref}
\clearpage
\pagebreak
\newpage

\widetext
\begin{center}
\textbf{\large Supplementary material for ``Spectral purification of microwave signals with disciplined dissipative Kerr solitons"}
\end{center}

\section{Locking range measurement with PM injection locking}

\begin{figure} [h]
\centerline{\includegraphics[width=0.6\columnwidth]{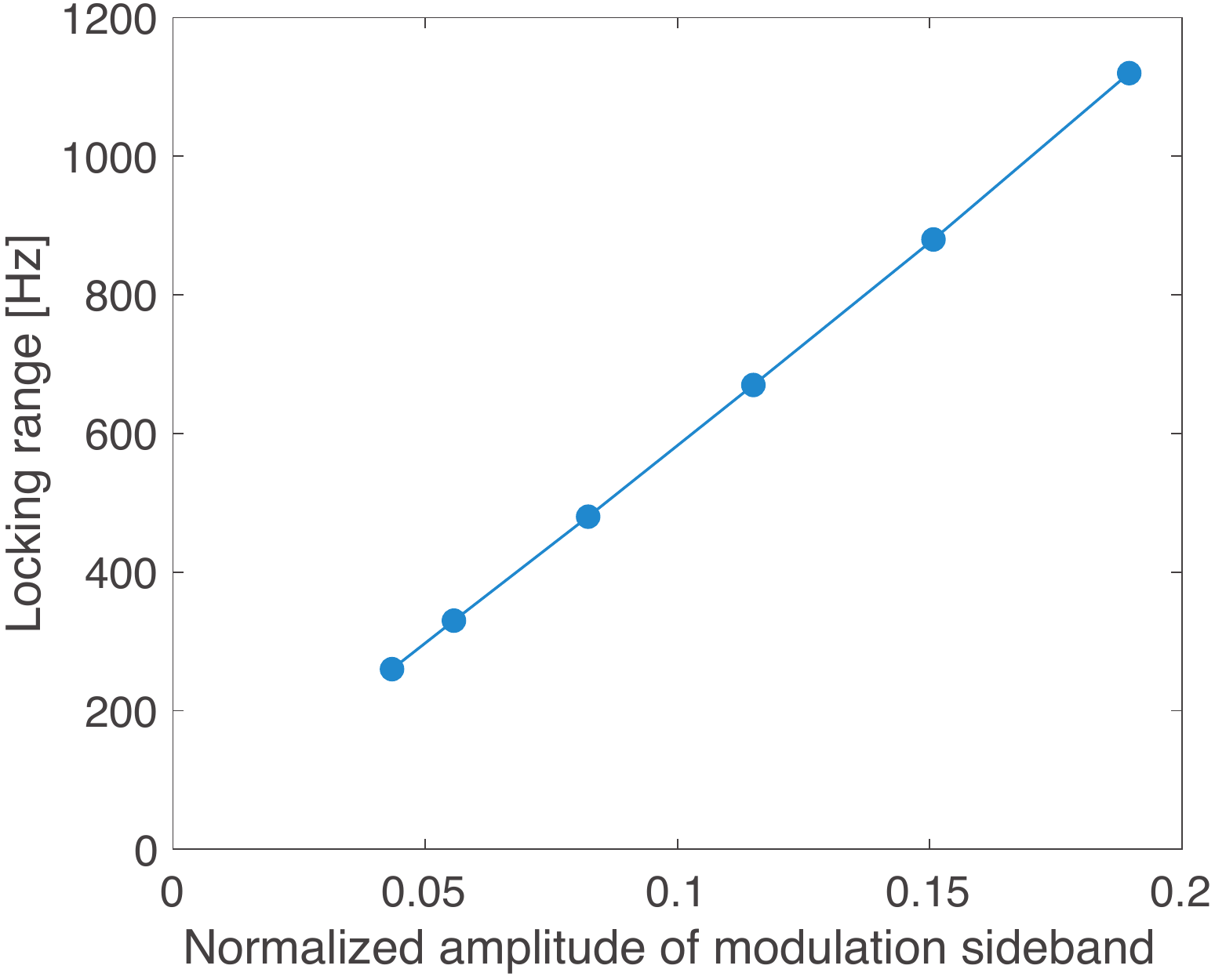}}
\caption{Measured locking ranges with laser-resonance detuning of 11.8\,MHz. The normalized amplitude of modulation sideband corresponds to $|\frac{\epsilon}{2}|$, where $\epsilon$ is defined in Eq.\,1 in the main text.}
\label{PM}
\end{figure}

In the main text we show the measured locking ranges with AM applied to the pump laser. We also measure the locking ranges with PM scheme and with varied PM strength, which are presented in Fig.\,\ref{PM}. We note that the locking range can be influenced by mode coupling between different mode families (see the next section), and that the resulting effect on the locking range can vary with temperature and fiber-resonator coupling condition. However, the results show the similar linear dependence of the locking range on the modulation strength with relatively small PM strength.

\section{Mode crossings' effect on the locking range}

\begin{figure} [h]
\centerline{\includegraphics[width=0.6\columnwidth]{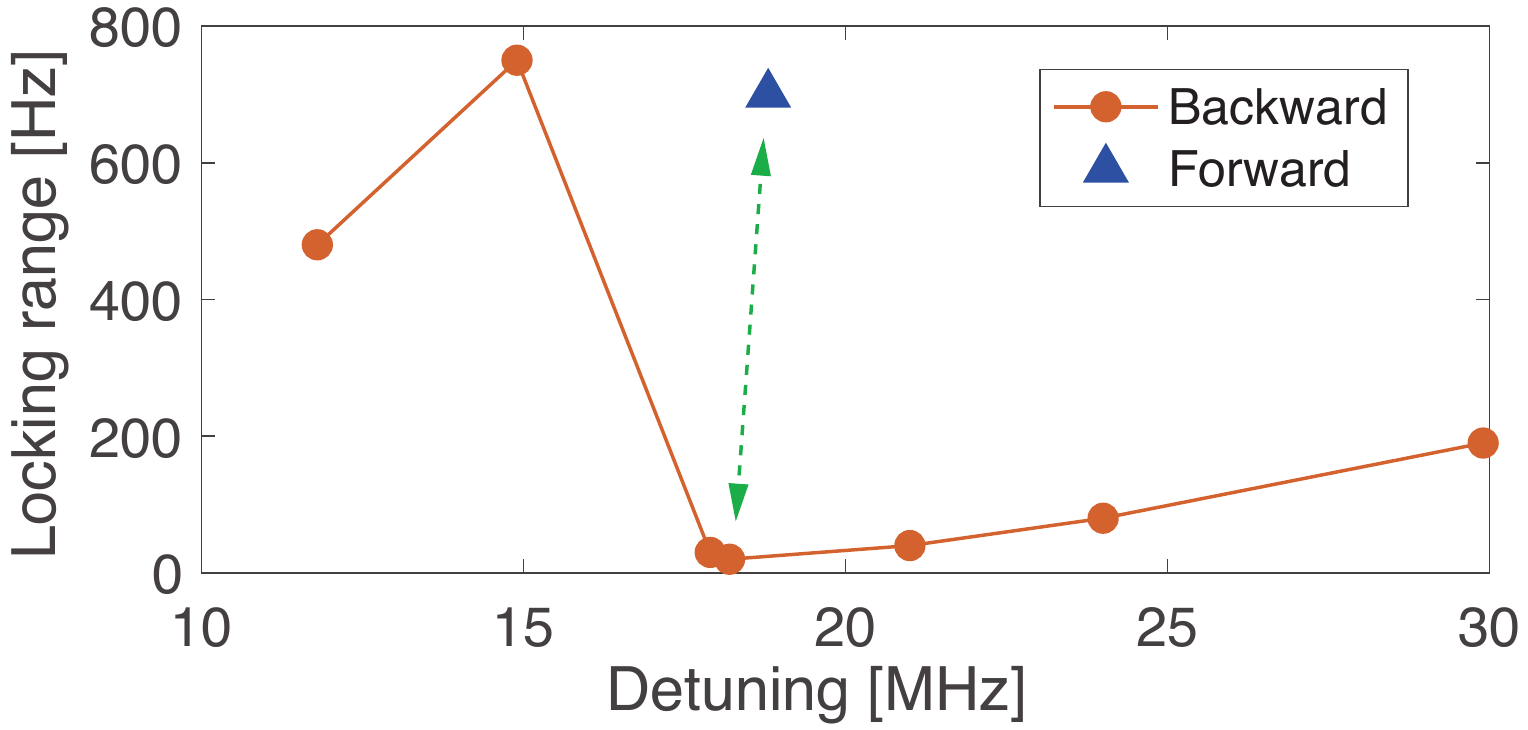}}
\caption{Shift of the DKS $f_\text{rep}$ as the laser-resonance detuning is tuned forward (to high frequency) and backward (to low frequency) respectively. Four comb spectra corresponding to the four points labeled on the curves are presented, showing that the strong bistability of $f_\text{rep}$ shift is caused by the mode coupling between different mode families, which also causes the appearance and disappearance of the single-mode dispersive wave indicated by the red arrows.}
\label{ranges}
\end{figure}

We apply PM injection locking by modulating the pump laser with frequencies close to the cavity FSR ($\sim$14.09\,GHz) with an EOM. We measure the locking ranges at different laser-resonance detunings but with the same PM strength. For the measurement the laser-resonance detuning is tuned from high frequency (30\,MHz) to low frequency (11.8\,MHz), which we referred to as backward tuning. We observe that the locking ranges change dramatically and erratically as the detuning changes, as presented in Fig.\,\ref{ranges}. Then we tune the detuning back to high frequency (forward tuning) and measure the locking range at detuning of 18.8\,MHz, which is also presented in the same figure. From this comparison we observe strong bistability of the locking range at the almost same detuned positions, as indicated by the green double-arrow.

\begin{figure} [h]
\centerline{\includegraphics[width=0.6\columnwidth]{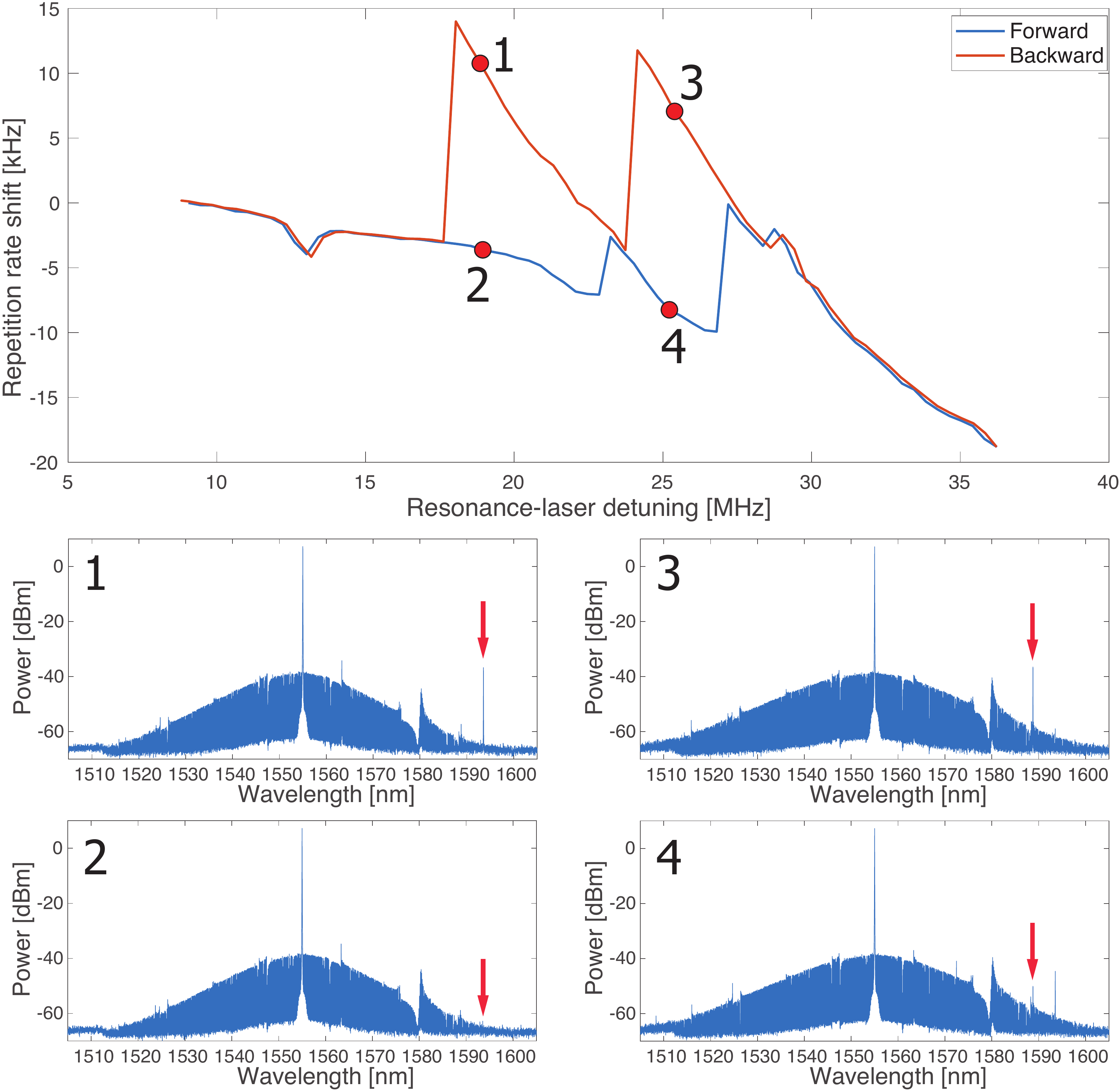}}
\caption{Shift of the DKS $f_\text{rep}$ as the laser-resonance detuning is tuned forward (to high frequency) and backward (to low frequency) respectively. Four comb spectra corresponding to the four points labeled on the curves are presented, showing that the strong bistability of $f_\text{rep}$ shift is caused by the mode coupling between different mode families, which also causes the appearance and disappearance of the single-mode dispersive wave indicated by the red arrows.}
\label{amx}
\end{figure}

We suspect that such behavior is induced by the mode coupling between different mode families \cite{guo2017intermode}, which can produce strong single-mode dispersive waves \cite{yi17single} that can modulate the CW background and trap the solitons \cite{wang2017universal} and consequently nullify the injection locking. To verify our speculation, we measure the DKS $f_\text{rep}$ and the comb spectrum continuously as we tune the detuning backward and then forward. Fig.\,\ref{amx} shows the experimental results. Indeed, bistability of the repetition rate shows with different tuning directions, due to the appearance and disappearance of the single-mode dispersive waves that can shift $f_\text{rep}$ through soliton recoil \cite{PhysRevA.95.043822}.

\section{Numerical simulation of PM-to-PM transfer function}
\begin{figure} [h]
\centerline{\includegraphics[width=0.6\columnwidth]{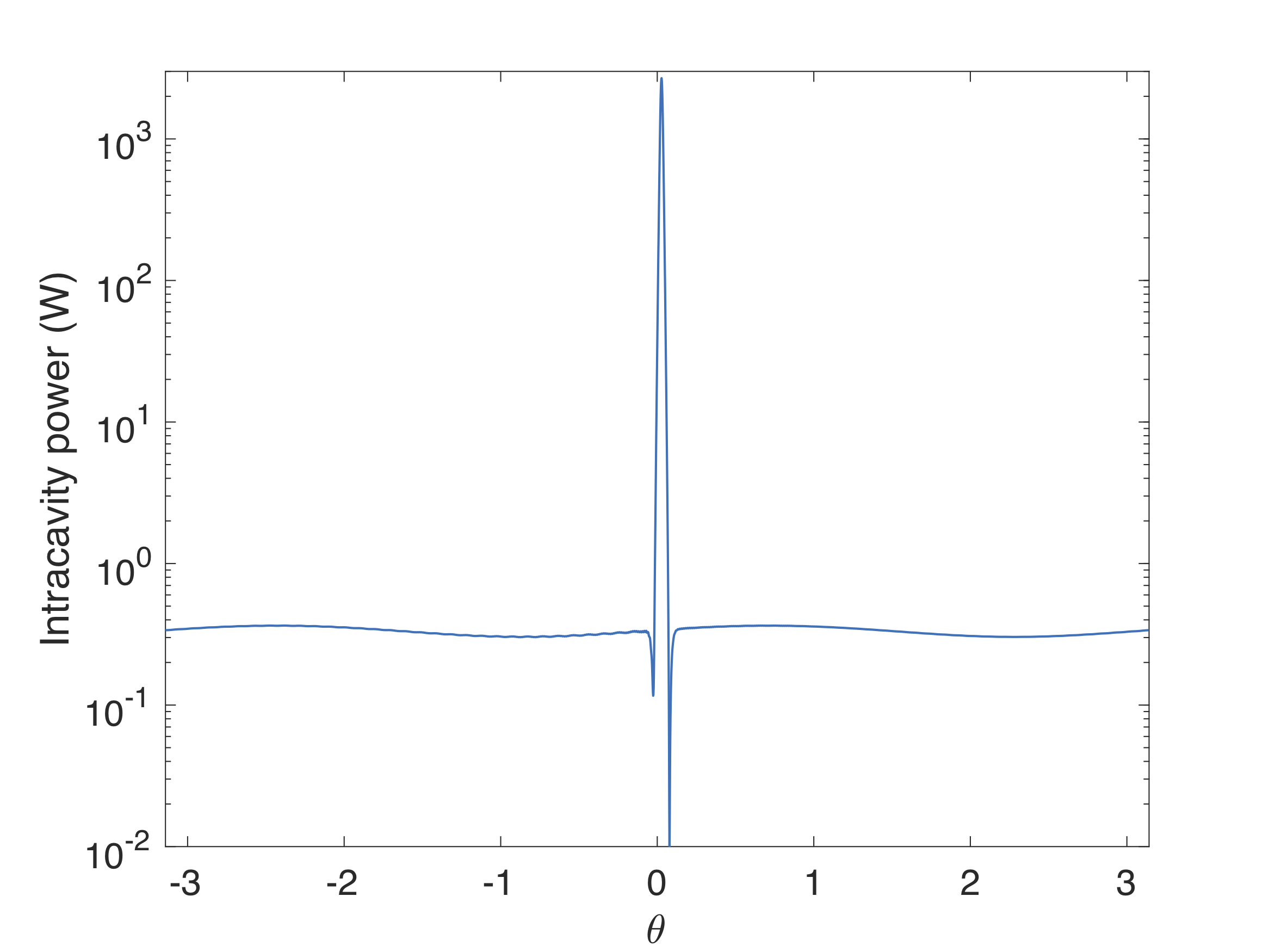}}
\caption{Temporal profile of a soliton that is trapped by the modulated intracavity background. The pump power is 200\,mW. Phase modulation with $\epsilon=0.32$ (5\,mW for the modulation sideband power) is used for the simulation.}
\label{fig10}
\end{figure}

LLE-like model described in \cite{guo2017intermode} is used to perform the simulation with adaptive Runge-Kutta method. We include both the second-order dispersion ($D_2/2 \pi = 2$\,kHz) and third-order dispersion ($D_3/2 \pi = -4$\,Hz) in $D_{\rm{int}}(\mu)$. Due to the high-order dispersion effect, the soliton repetition rate displays a frequency shift of $\sim-7.3$\,kHz with respect to the cavity FSR (14\,GHz). We set the input microwave signal ($\Omega$) to match the repetition rate so the solitons can be trapped by the modulated intracavity background. The central pump power is 200\,mW and the PM sideband power is set to be 10\,mW, which corresponds to $\frac{\epsilon^2}{4}=0.05$. After tens of photon decay time the position of the soliton with respect to the modulated background does not change anymore, indicating that $f_\text{rep}$ is injection-locked. In Fig.\,\ref{fig10} we plot the trapped soliton and the modulated intracavity background.

\begin{figure} [h]
\centerline{\includegraphics[width=0.6\columnwidth]{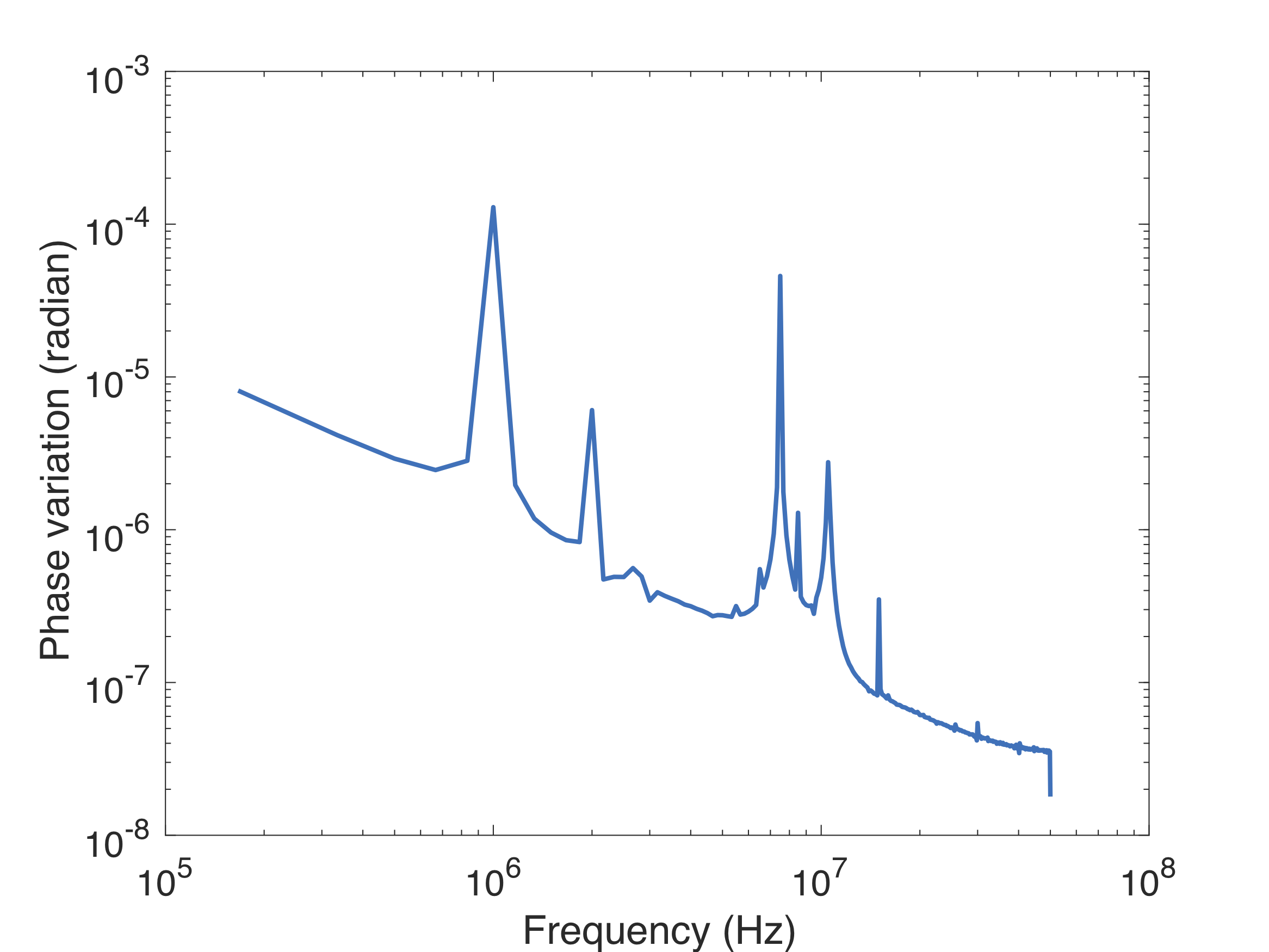}}
\caption{Spectrum of the phase variation of the injection-locked $f_\text{rep}$. The PM amplitude (phase deviation) on the optical sidebands is 0.1\,radian. The PM frequency is 1\,MHz.}
\label{fig20}
\end{figure}

After the soliton is trapped, we apply PM on the injected microwave signal. This modulation is directly converted into PM on the optical modulation sidebands. We choose the modulation amplitude (PM deviation) to be 0.1\,radian and let the simulation run for 8 -- 20 modulation periods. The phase response of the disciplined soliton repetition rate is derived with Eq.\,2 in the main text. Then we apply fast Fourier transform (FFT) on the phase response in the time domain to obtain the spectrum of the phase variation \cite{matsko2015noise}. Fig.\,\ref{fig20} shows the spectrum of the phase variation of the injection-locked $f_\text{rep}$ as PM at 1\,MHz is applied to the optical modulation sidebands.

\begin{figure} [h]
\centerline{\includegraphics[width=0.6\columnwidth]{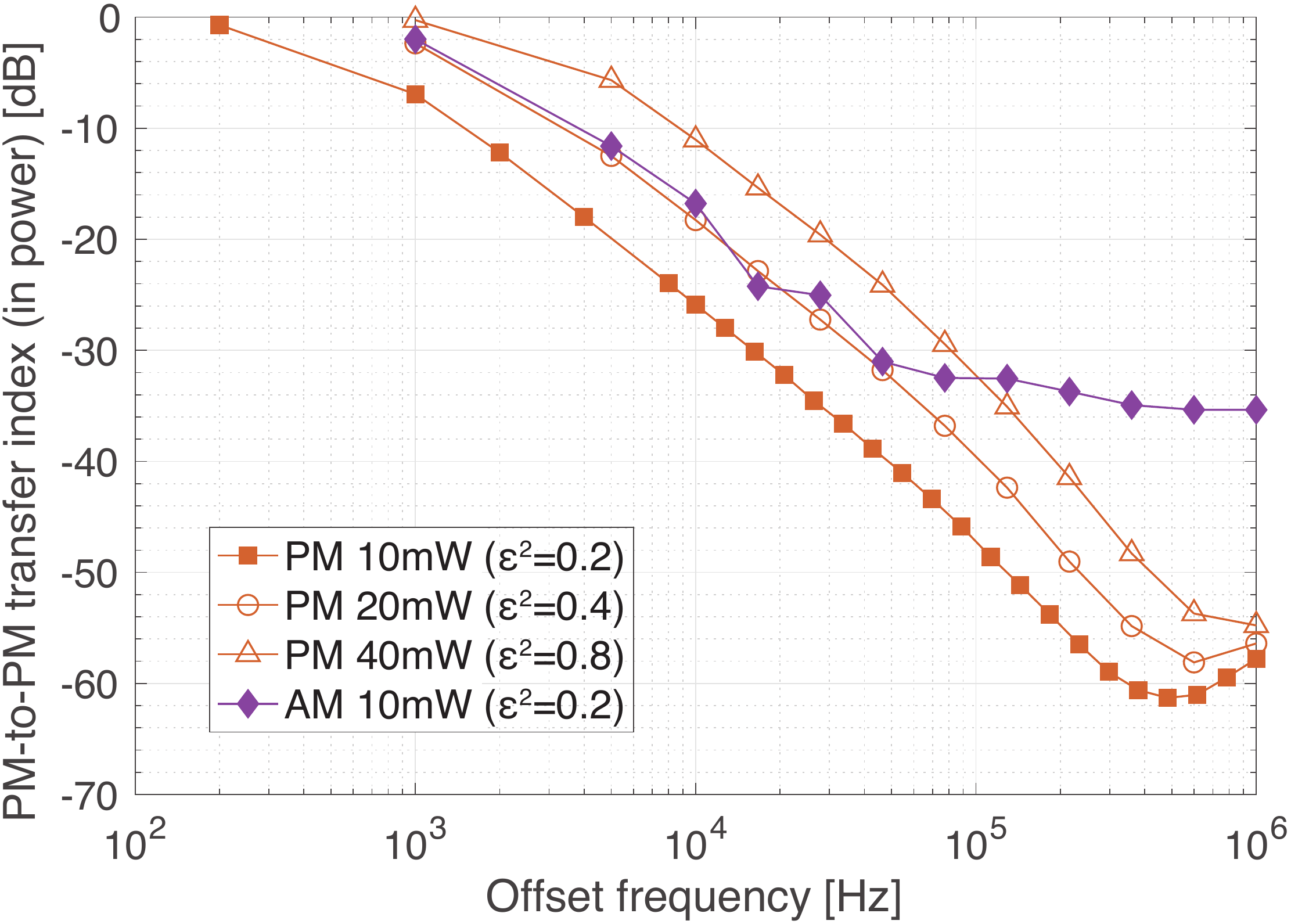}}
\caption{Simulated PM-to-PM transfer functions with different injection-locking schemes and varied laser modulation strengths. The optical sideband powers are indicated in the legend.}
\label{fig30}
\end{figure}

With the phase response spectrum we can calculate the PM-to-PM transfer indices at particular PM frequencies. We choose discrete PM frequencies from 200\,Hz to 1\,MHz to simulate the transfer function shown in the main text. For comparison, we also carry out simulations of PM injection locking with larger $\epsilon$ and AM injection locking. The simulated results are summarized in Fig.\,\ref{fig30}. As the modulation strength increases, the transfer function amplitude rises, showing that the soliton trapping bandwidth is enhanced. Moreover, AM injection locking exhibits higher transfer indices at offset frequencies higher than $\sim100$\,kHz.

\section{Experimental measurement of PM-to-PM transfer function}

\begin{figure}
\centerline{\includegraphics[width=0.6\columnwidth]{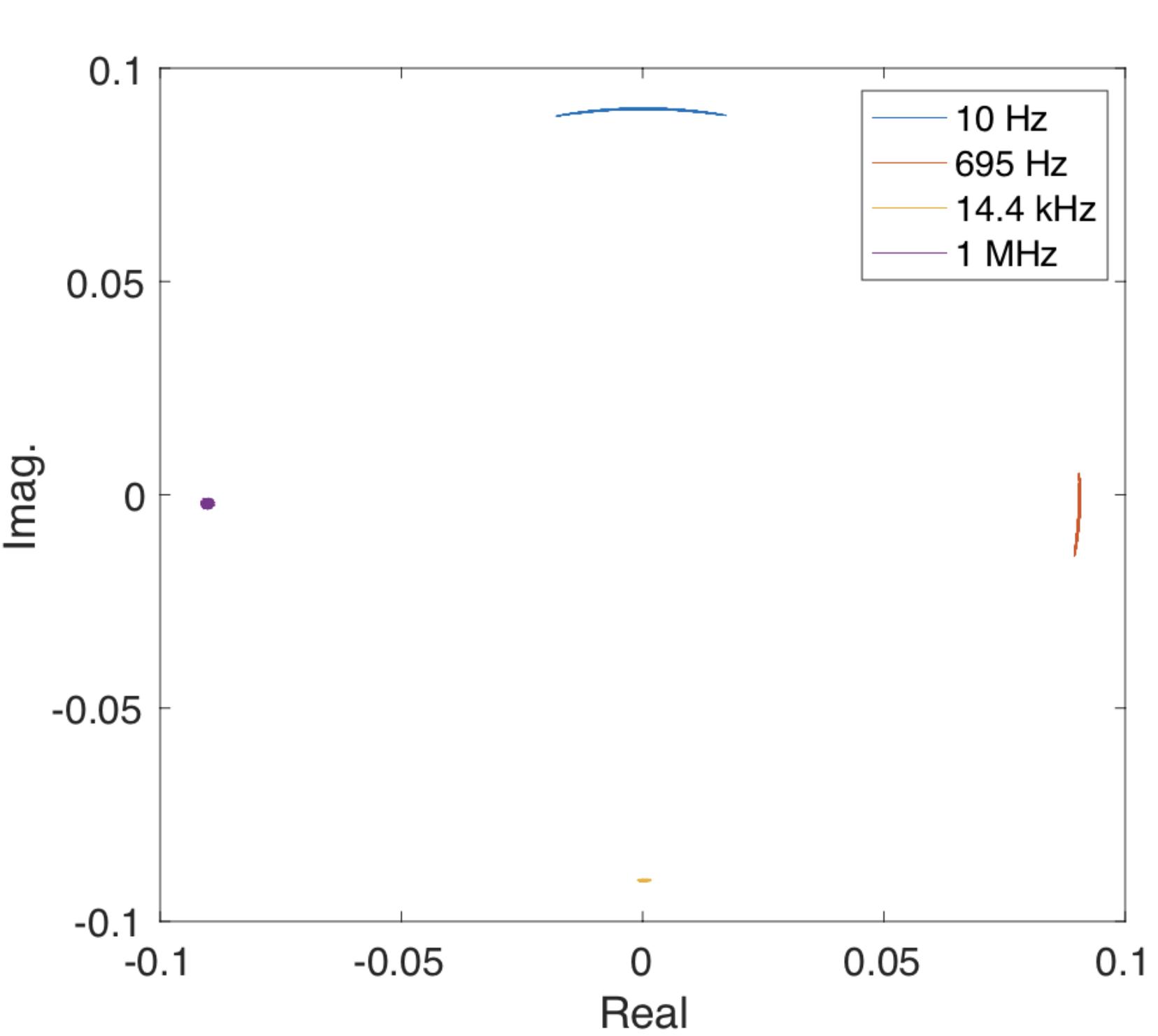}}
\caption{Measured results of the microwave signal from the IQ test. The amplitude and the phase responses to the injected PM on the input microwave signal can be derived from the data in complex plane. The 4 sets of data are corresponding to 4 different PM frequencies that are indicated in the figure. The phase deviation of the PM on the input microwave signal is set to be 0.1\,radian.}
\label{fig40}
\end{figure}

To verify the simulated PM-to-PM noise transfer function, we implement PM on the input microwave signals via internal function of the synthesizer that is used to drive the optical phase modulator. The PM amplitude is set to be 0.1\,radian, 1\,radian and 2\,radian respectively. After the soliton repetition rate is injection locked, the PM on the microwave signals is activated, and the in-phase-and-quadrature (IQ) demodulation function of the spectrum analyzer is utilized to measure the phase and amplitude responses of the microwave signals generated by the fast photodetector that detects the DKS repetition rate. The demodulation frequency on the analyzer is set equal to the synthesizer frequency and both devices are referenced to a common clock oscillator. In Fig.\,\ref{fig40} we plot the measured results of the IQ phasors at 4 different PM frequencies shown in the complex plane when the PM phase deviation is 0.1\,radian. We note that the approach can also be used to measure the PM-to-AM noise transfer function.

\begin{figure}
\centerline{\includegraphics[width=0.6\columnwidth]{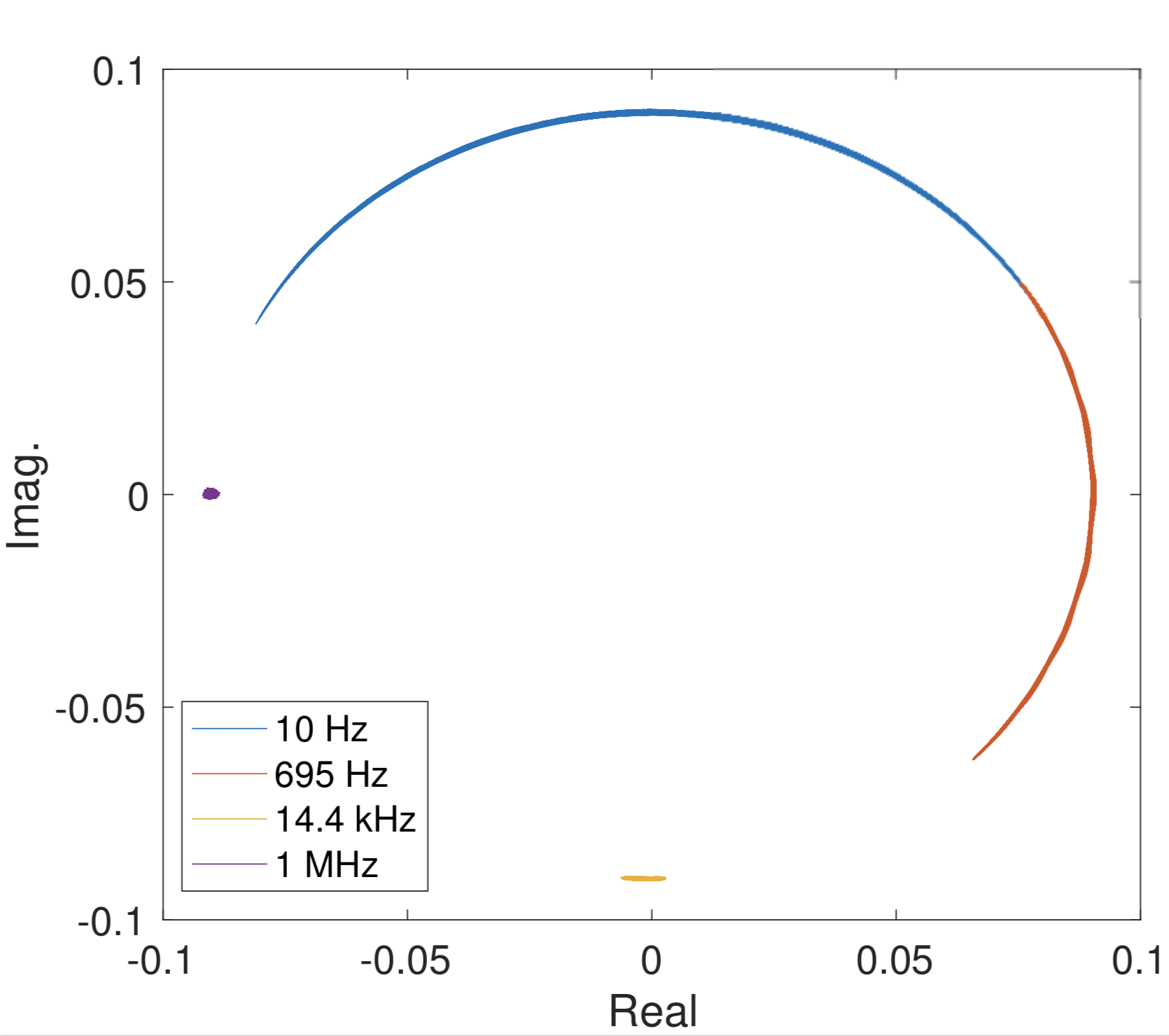}}
\caption{Measured results of the microwave signal from the IQ test with phase deviation of 1\,radian.}
\label{fig50}
\end{figure}

As shown in the main text, as the PM frequency on the input microwave signal increases, the phase deviation of the purified signal decreases. To quantify this effect, the phase deviation is extracted from the IQ measurement and the transfer function is obtained as
$20 \operatorname{log}_{10}\left(\dfrac{\Delta\phi^{(\text{measured})}(\omega)}{\Delta\phi^{(\text{in})}(\omega)}\right)$. Eventually, the response amplitude is below the instrumental noise floor of the phase noise analyzer, which is evidently displayed as the flat floor on the transfer function at high frequencies. To increase the dynamical range of the measurement, the measurement is repeated with larger phase deviations (1\,radian and 2\,radians) on the modulation and re-take the IQ measurement. Fig.\,\ref{fig50} shows 4 sets of measured results at 4 different PM frequencies. We see that at very high PM frequency (1\,MHz) the response is limited to the same instrumental noise floor. However, because of the larger input phase deviation, the noise floor level of the transfer function decreases, as shown by Fig.\,5 in the main text.

\end{document}